\documentclass[preprint,
 amsmath,amssymb,
 aps, 
pra,
]{revtex4-2}

\usepackage{graphicx}
\usepackage{dcolumn}
\usepackage{bm}

\usepackage[utf8]{inputenc}
\usepackage[english]{babel}
\usepackage{amsmath}
\usepackage{amsthm}
\usepackage{amsfonts}
\usepackage{amssymb}
\usepackage{graphicx}
\usepackage{booktabs}
\usepackage[hidelinks]{hyperref}
\usepackage{multirow}
\usepackage{tikz}
\usepackage{verbatim}
\usepackage[ruled,vlined]{algorithm2e}
\usepackage{enumitem}
\usepackage{float}
\usepackage{color}
\usepackage[normalem]{ulem}
\usepackage{siunitx} 
\usepackage{isotope} 
\usepackage{soul}
\usepackage{makecell}

\def\R {{\mathbb R}}

\newcommand{\zd}{\delta}

\newcommand{\zs}{\sigma}

\newcommand{\ze}{\varepsilon}

\newcommand{\zg}{\gamma}

\newcommand{\zh}{\eta}
\newcommand{\zn}{\nu}

\newcommand{\dif}{\; \textrm d }

\newcommand{\dsp}[2]{\frac{  \partial^2 #1 }{  \partial #2^2   } }
\newcommand{\dst}[2]{\frac{  \dif^2 #1 }{  \dif #2^2   } }
\newcommand{\dpp}[2]{\frac{  \partial #1 }{  \partial #2   } }
\newcommand{\dpt}[2]{\frac{  \dif #1 }{  \dif #2   } }

\begin{document}


\title{\textbf{Optimal control in phase-space applied to minimal-time transfer of thermal atoms in optical traps}}

\author{Omar Morandi}
 \affiliation{Department of Mathematics and Informatics "Ulisse Dini", University of Florence, Viale Morgagni 67/A, 50134 Florence, Italy}
            
\author{Sara Nicoletti}
 \email{Contact author: sara.nicoletti@unifi.it}
\affiliation{Department of Mathematics and Informatics "Ulisse Dini", University of Florence, Viale Morgagni 67/A, 50134 Florence, Italy}

\author{Vladislav Gavryusev}
\affiliation{Department of Physics and Astronomy, University of Florence, Via G. Sansone 1, 50019 Sesto Fiorentino, Italy}
\affiliation{European Laboratory for Non-Linear Spectroscopy, University of Florence, Via N. Carrara 1, 50019 Sesto Fiorentino, Italy}
\affiliation{National Institute of Optics, National Research Council, Via N. Carrara 1, 50019 Sesto Fiorentino, Italy}
\author{Leonardo Fallani}
\affiliation{Department of Physics and Astronomy, University of Florence, Via G. Sansone 1, 50019 Sesto Fiorentino, Italy}
\affiliation{European Laboratory for Non-Linear Spectroscopy, University of Florence, Via N. Carrara 1, 50019, Sesto Fiorentino, Italy}
\affiliation{National Institute of Optics, National Research Council, Via N. Carrara 1, 50019 Sesto Fiorentino, Italy}



\begin{abstract}
We present an optimal control procedure for the non adiabatic transport of ultracold neutral thermal atoms in optical tweezers arranged in a one-dimensional array, with the focus on reaching a minimal transfer time. The particle dynamics is modeled first using a classical approach through the Liouville equation and second through the quantum Wigner equation to include quantum effects. Both methods account for typical experimental noise described as stochastic effects through Fokker-Planck terms. The optimal control process is initialized with a trajectory computed for a single classical particle and determines the phase-space path that minimizes transport time and ensures high transport fidelity to the target trap. This approach provides a fast and efficient method for relocating atoms from an initial configuration to a desired target arrangement, minimizing time and energy costs while ensuring high fidelity. Such an approach may be highly valuable to initialize large atom arrays for quantum simulation or computation experiments. 
\end{abstract}

\maketitle


\section{\label{sec:level1}Introduction}

Steering a quantum system from an initial state to a target state is of outmost relevance in quantum information science~\cite{Nielsen2010}. In particular, engineering methods for the high-fidelity preparation of quantum states plays a central role in the development of quantum technologies, where physical systems are being used to develop new sensors, simulators, computers or communication devices~\cite{Koch2022}. In this context, implementations based on programmable arrays of trapped neutral atoms have emerged as a highly controllable platform for the realization of quantum simulators and quantum computers, as well as for the exploitation of quantum entanglement in sensing and metrology~\cite{Saffman2019,Henriet2020,Bluvstein2023}. In this platform the atoms are trapped in micron-scaled optical traps, called optical tweezers~\cite{Gieseler2021}, obtained by tightly focusing individual laser beams through a high-resolution microscope~\cite{Frese2000,Schlosser2001,Schlosser2002} or by employing spatial light modulators~\cite{Kim2016,Barredo2018}. The position and intensity of the tweezers can be individually controlled, leading to the realization of arrays of traps with custom and reconfigurable geometry. Due to light-assisted collisions~\cite{Gruenzweig2010,Sompet2013,Carpentier2013,Fung2015,Fung2016}, it is possible to trap a single atom in each tweezer, isolating single-particle quantum systems that can be individually initialized, steered, and measured via direct imaging. In several applications, e.g. for the realization of quantum simulators and quantum processors, atom-atom interactions can then be activated at will by exciting the atoms towards high-lying Rydberg states~\cite{Barredo2014,Barredo2016,Browaeys2016,Gavryusev2016,Norcia2018,FerreiraCao2020,Browaeys2020,Wilson2022}. 
    
Although methods for quasi deterministic loading are being developed~\cite{Gruenzweig2010,Lester2015,Fung2015,Brown2019}, the process of trapping atoms in the tweezers from a finite-temperature laser-cooled atomic cloud is inherently stochastic. This leads to a random occupation of the traps, which requires the implementation of rearrangement protocols where atoms are transported from the initial to the target position, to create an ordered array with zero configurational entropy~\cite{Beugnon2007,Miroshnychenko2006,Schlosser2012,Nogrette2014,Kim2016,Endres2016,Barredo2018,Cooper2018,OhldeMello2019,Schymik2020,Glicenstein2021,Schymik2022}. Typically, such transport is performed in an adiabatic way~\cite{Lengwenus2010,Schlosser2012}, that is, on a timescale (typically on the order of milliseconds) that is much longer than the timescale of motion in the individual traps (typically on the order of $\qty{10}{\micro\second}$), avoiding excitation of atoms during the transport through parametric heating~\cite{Savard1997,Boulier2019}. However, as the array is scaled to hundreds or thousands of atoms, this strategy poses severe constraints on the time needed for the rearrangement process, which is detrimental both because it reduces the duty cycle of the experiment and because it can lead to an increased infidelity of the state preparation due to the finite lifetime of the atoms in the traps.

Because of this critical issue, it is desirable to implement nonadiabatic protocols~\cite{Couvert2008,Murphy2009,Chen2011,Torrontegui2011,Negretti2013,GueryOdelin2019}, where the transport is performed on much shorter timescales, which can be comparable to that of the atom dynamics in the trap. The idea is to employ engineered trajectories where the atom evolves through excited motional states during the trap motion, ending up (almost) at rest in the final trap position. Strategies of non adiabatic transport have already been investigated in some experimental works.
In Ref.~\cite{Muldoon2012} the authors implemented a method through a digital micromirror device to dynamically control the position of neutral atoms in optical tweezers by using the "release and recapture" method with minimal heating, reaching a fidelity of $55\%$, while in~\cite{Kim2016,Lee2016} the atoms were concurrently rearranged using a spatial light modulator (SLM) with a fidelity of $86\%$ and recent advances using SLMs with kilohertz update rates~\cite{Lin2024,Knottnerus2025} have reached a fidelity above $99\%$. In Ref.~\cite{Hwang2023} the authors provided and experimentally demonstrated freely flying atoms thrown and caught by optical tweezers realized using an acousto-optical deflector (AOD)~\cite{Duocastella2020,Ricci2024} with a transport efficiency of $94(3)\%$. Various methods to engineer nonadiabatic transport protocols have been considered in previous theoretical works.
The problem of transporting a quantum state while maintaining the encoded quantum information was explored in Ref.~\cite{Murphy2009}, considering a moving harmonic potential well and allowing for imperfect control over the system to model experimental limitations. Optimal control theory paired with an invariant-based method was applied in Ref.~\cite{Chen2011} to identify solutions that minimize either motion time, displacement or transient energy, considering also anharmonic traps and noise sources~\cite{Torrontegui2011}. In Ref.~\cite{Pagano2024} the authors studied the application of a quantum optimal control procedure to transport neutral atoms in optical tweezers using the dressed chopped random basis (dCRAB) method~\cite{Mueller2022}. Shortcuts to adiabaticity (STA)~\cite{GueryOdelin2019} were considered in~\cite{Hwang2025,Cicali2024} as an efficient alternative for fast and reliable atom transport control. Finally, recent experiments~\cite{Bluvstein2022,Bluvstein2023} have demonstrated the capability to transport atoms in parallel between several positions while preserving the quantum state coherence in order to realize quantum gates, perform error correction, and read out the final state.

The transport of ultracold trapped atoms has been extensively studied in recent years, with particular emphasis on optimization protocols and efficient high-fidelity transfer.
Several approaches have been explored for trapped ions, with a focus on strategies that minimize the vibrational excitation and the  decoherence during the transfer process. Transport optimization has been implemented in ion traps to obtain fast shuttling with minimal heating effects~\cite{Huber2008,Bowler2012,Walther2012}. Optimal transport has also found several applications in ultracold atoms in optical dipole traps~\cite{Couvert2008,Chen2010} and lattices, including Bose-Einstein condensates (BECs). Experimental studies have demonstrated the possibility to achieve fast transport of single atoms while preserving quantum coherence, reducing motional excitations, and optimizing loading efficiency in optical lattices~\cite{Rosi2013}. In the context of BECs, non adiabatic transport of condensates has been explored, revealing strategies to mitigate excitations induced by trap movement~\cite{Jaeger2014}. Theoretical models have analyzed the role of anharmonic potentials and optimized control protocols are now available to enhance transport fidelity and suppress unwanted excitations~\cite{Buecker2013}. Optimal control techniques have been proposed to achieve fast, high-fidelity transport of BECs, ensuring minimal energy cost and robustness against experimental imperfections~\cite{Hohenester2007}.

In this work, we derive an optimal control procedure aimed to steer neutral atoms with optical tweezers.
For the sake of simplicity, we assume that the system dynamics is restricted to one dimension, with the tweezer moving in a horizontal line. This assumption is quite realistic as most of the experiments are performed either on one-dimensional (1D) linear arrays or in 2D arrays being assembled with a set of linear displacements~\cite{Endres2016,Glicenstein2021}. The control parameters are obtained by minimizing the distance of the particle distribution function within a target region of the phase-space and at the same time maintaining the energy cost for the control as small as possible, in other words by reducing its $L^2$ norm. 
In particular, we consider two static traps, one located at the initial atom position and the other at the target position, and a movable tweezer potential on which we act to reach the target position, as in Ref.~\cite{Barredo2016}. The controlled quantities are the depth and the center coordinates of the tweezer and we pair them with a cost functional that should be kept as small as possible. First, the model is formulated in terms of a classical deterministic particle obeying the Hamiltonian equations. Then we study the statistical ensemble control problem governed by the Liouville Fokker-Planck equation, where any source of noise is modeled as an external thermal bath with a certain temperature. Finally, we consider a fully quantum problem, studying an ensemble control problem governed by the Wigner equation integrated with the Fokker-Planck terms to consider stochastic effects.

The paper is organized as follows. In Sec.~\ref{sec:2} we derive the optimal trajectories for a single classical particle. In Sec.~\ref{sec:3} we introduce the statistical description for the classical dynamics of an atom ensemble. In Sec.~\ref{sec:4} we stress the model to take into account quantum effects and we consider a formulation for the atomic distribution based on the Wigner equation. In Sec.~\ref{sec:5} we highlight the advantages of our method and we compare it with different approaches. In Sec.~\ref{sec:6} we summarize the work and discuss our conclusions.

\section{Model}
\label{sec:2}
We study the optimal 1D transport of neutral atoms between two static optical traps, located at the initial position $A$ and final position (target) $B$, respectively. The goal of our optimal control procedure is to design the time evolution of an additional moving optical tweezer field, which should steer the atoms from $A$ to $B$ in the minimum possible amount of time (see Fig.~\ref{fig_Opt_traj_02}). The optical fields of both the moving tweezer and the static traps are modeled by a parametrized Gaussian function~\cite{Endres2016}
\begin{align}
	U_C = v(t) \, e^{-(x-u(t))^2/\zs^2_x} \; ,\label{tweez_pot}
\end{align} 	
where $u(t)$ and $v(t)$ are the center position and the amplitude of the trapping potential, respectively, while the tweezer beam size $\zs_x$ is kept constant. In Refs.~\cite{Torrontegui2012,Zhang2015,Zhang2016,Amri2019} a parabolic approximation was used. The time-dependent quantities $u(t)$ and $v(t)$ are the two control parameters of the tweezers. The movable trap is suddenly turned on at the initial time $t=0$ and turned off once the final time $t_f$ is reached.
	
The transfer of the atom should be performed in a optimal way according to the following criteria. We design the temporal profile of the control parameters $u(t)$ and $v(t)$ and the interval $[0,t_f]$ during which the control procedure should be achieved, by maximizing the success probability of the protocol, i.e., the probability that at the final time $t_f$ the atom is found at rest inside the target trap $B$, and concurrently minimizing the energetic cost of the control and of the final time $t_f$. 

To measure the energy cost associated with the control, we define the following functional, which should then be maintained as small as possible:  
\begin{equation}
\label{eq:cost}
k(u,v)=\frac{1}{2}\int_0^{t_f}\biggl[\gamma_u|u(t)|^2+\gamma_v|v(t)|^2 + \nu_u\biggl|\dpt{u}{t}\biggr|^2+\nu_v\biggl|\dpt{v}{t}\biggr|^2\biggr]\dif t \; ,
\end{equation}
Here $\gamma_u,\gamma_v,\nu_u,\nu_v>0$ are parameters that can be used to tailor the trade-off between the energetic cost magnitude over the success rate of the protocol. The integral terms in the cost functional~\eqref{eq:cost} measure the energy cost associated with the control and represent the $H^1$-norm of the control signal. In particular, the terms containing time derivatives penalize fast oscillating solutions over slow regular solutions and ensure that the control parameter profiles resulting from the optimal control procedure vary smoothly in time.  
		
We describe the optimal control of the atom trajectories driven by the optical tweezer field at three increasing degrees of precision. First, we assume that the particle dynamics is completely deterministic. We consider each atom to be well described by a single classical trajectory with known initial position that evolves in the absence of any source of external noise. This idealized description of the atom dynamics allows us to calculate the optimal trajectory efficiently.

In the second step, we introduce in our model the main sources of uncertainties found in the experimental manipulation protocols. The initial condition of the atom is known only approximately and various sources of external perturbations are typically present.
Common experimental sources of random fluctuations~\cite{Bernien2017,Sheng2021,Bluvstein2022,Bluvstein2023} are the beam pointing and depth fluctuations of the tweezer trap~\cite{Schymik2022}, leading to parametric heating~\cite{Savard1997} and limiting the atom lifetime, the finite frequency bandwidth of the trap-position driving signal, and waist fluctuations along the path due to optical imperfections, laser phase noise, finite atom temperature, and heating due to collisions with other atoms that may be too close to the motion trajectory. These effects can be described together as a stochastic Markovian process that leads to an effective broadening of the initial statistical distribution of atoms in the position and momentum space and with the interaction with a thermal bath during the transport. For this reason, we describe the atom system in terms of a classical ensemble of particles described by a statistical distribution in the phase-space. Dissipation effects and external noise sources are modeled by a Liouville  Fokker-Planck  equation for the particle density. We remark that at this stage the atom dynamics is purely classical.
	
Finally, we describe the atom dynamics in a fully quantum context. In order to highlight the correction to the previous classical results, we adopt a kinetic description of the quantum motion provided by the Wigner formalism of the pseudo distribution function. The external noise is still modeled in terms of Fokker-Planck terms included in the equation. 
Due to the non linearity of the optimality systems associated with the optimal control, each step of our modelization of the controlled dynamics is used in the following one as a convenient initial guess to initialize the minimization procedure.    
	
\subsection{Deterministic transport}
The simplest way to describe the motion of a single atom of mass $m$, steered by the tweezers field, is to model the atom as a classical particle whose initial conditions are exactly known. 
With this assumption, the particle trajectory can be obtained by solving the Hamiltonian equations 
\begin{align} 
	& \dpt{}{t} \left(
	\begin{array}{c}
		x  \\
		p 
	\end{array}\right) =\left(\begin{array}{c}
        	\frac{p}{m}  \\[2mm]
	        -\dpp{U}{x} 
	      \end{array}\right)\;,\label{Ham_traj}
\end{align}
with initial conditions $x(0)=x_A$ and $p(0)=0$, where $x$ and $p$ denote the particle position and momentum, respectively. The total potential $U(u,v,x)=U_C(u,v,x)+U_A(x)+U_B(x)$ is given by the sum of the potentials associated with the moving tweezer ($U_C$) and with the initial ($U_A$) and target ($U_B$) traps. The shape of the static trap potentials is similar to the tweezer profile given in Eq.~\eqref{tweez_pot} and is given by $U_i =  U_0 \, e^{-[(x-x_i)/\zs_x)]^2}$, with $i=A,B$, where $U_0$ is the trap depth and $x_A$ and $x_B$ are the position of the initial and target traps, respectively. 
In the spirit of the optimal control procedure, we quantify the error associated with the control of the particle trajectory through the squared distance of the final position of the atom from the target position $x_B$ and of the final value of the momentum. We define the control error as $\Phi = \frac{\nu_x}{2}[x(t_f)-x_B]^2+\frac{\nu_p}{2} p(t_f)^2$, where $\nu_x$ and $\nu_p$ represent the weight parameters associated with the relevance of achieving the target position and momentum. The optimal control problem consists in determining the total flying time $t_f$ and the time-dependent value of the tweezer parameters $u$ and $v$ in the interval $[0,t_f]$, which is the solution of the constrained minimization problem   
\begin{align*}
    &\text{min}_{u,v,t_f} \left\{ \Phi +k +  \frac{\zn_{t_f}}{2} {t_f}^2\right\}\;  \\
    &  \textrm{s.t. Eq. }~\eqref{Ham_traj} \textrm{ holds true}\;,
\end{align*}
where $\nu_{t_f}$ is the weight associated with the parameter $t_f$. 
The set of nonlinear equations that provide the solution of the optimal problem is referred to as optimality conditions. A simple way to derive the optimality conditions is to express the optimal control problem in terms of a variational problem associated with a Lagrangian functional. We define the Lagrangian  
\begin{align}
    \label{cl_lagrangian}
	\mathcal{L}=&  \int_0^{t_f}\left[\left( \dot{x}- \frac{p}{m}\right){p}^h + \left( \dot{p}  +\partial_x U\right){x}^h \right] \dif t+ \Phi +k +  \frac{\zn_{t_f}}{2} {t_f}^2\;,
\end{align}
where we have introduced the Lagrangian multipliers $p^h$ and $x^h$, which are defined as adjoint variables in the optimal control framework. The adjoint variables constitute two additional unknowns of the problem.   

The necessary conditions for a set $(x,p,u,v,t_f)$ to be a solution of the minimization problem are obtained by imposing that the G\^{a}teaux derivatives of the Lagrangian functional with respect to all the parameters should vanish. 
We obtain the following equations for the adjoint variables:

\begin{align}
	\dot{p}_h=&x_h\frac{\partial^2 U}{\partial x^2}\label{Opt_adj_q},\\
	\dot{x}_h=&-\frac{p_h}{m}\label{Opt_adj_p}. 
\end{align}
Differing from the standard Cauchy problems, the values of the adjoint variables are known at the final time $t_f$. We have the final-value conditions $x_h(t_f)=-\nu_x p(t_f)$ and $p_h(t_f) =\nu_x [x_B -x(t_f)]$, calling that we assume $\nu_x=\nu_p$ as specified before. The control parameters $u$ and $v$ and the final time $t_f$ are obtained by solving the equations
\begin{align}
	\label{Opt_cont_u}
	&\nu_{u}  \dst{u}{t}-	\gamma_u u=-x_h\frac{\partial^2 U}{\partial x\partial u},\\
	&\nu_{v}  \dst{v}{t}-\gamma_v v=- x_h\frac{\partial^2 U}{\partial x\partial v},\\
	&\nu_{t_f} t_f=\biggr(-\frac{\nu_x}{m} [x_B-x(t_f)]+\nu_p\left.\frac{\partial U}{\partial x}\right|_{t=t_f}\biggl)p(t_f)-\frac{\gamma_u}{2}u^2(t_f)-\frac{\gamma_v}{2}v^2(t_f).	\label{Opt_cont_T}
\end{align}
In conclusion, the optimality system consists of Eq.~\eqref{Ham_traj} for the atom trajectory, Eqs.~\eqref{Opt_adj_q} and~\eqref{Opt_adj_p} for the adjoint trajectory, and Eqs.~\eqref{Opt_cont_u} and~\eqref{Opt_cont_T} for the control parameters. 

\begin{figure}[!ht]
	\centering
    \includegraphics[width=0.65\textwidth]{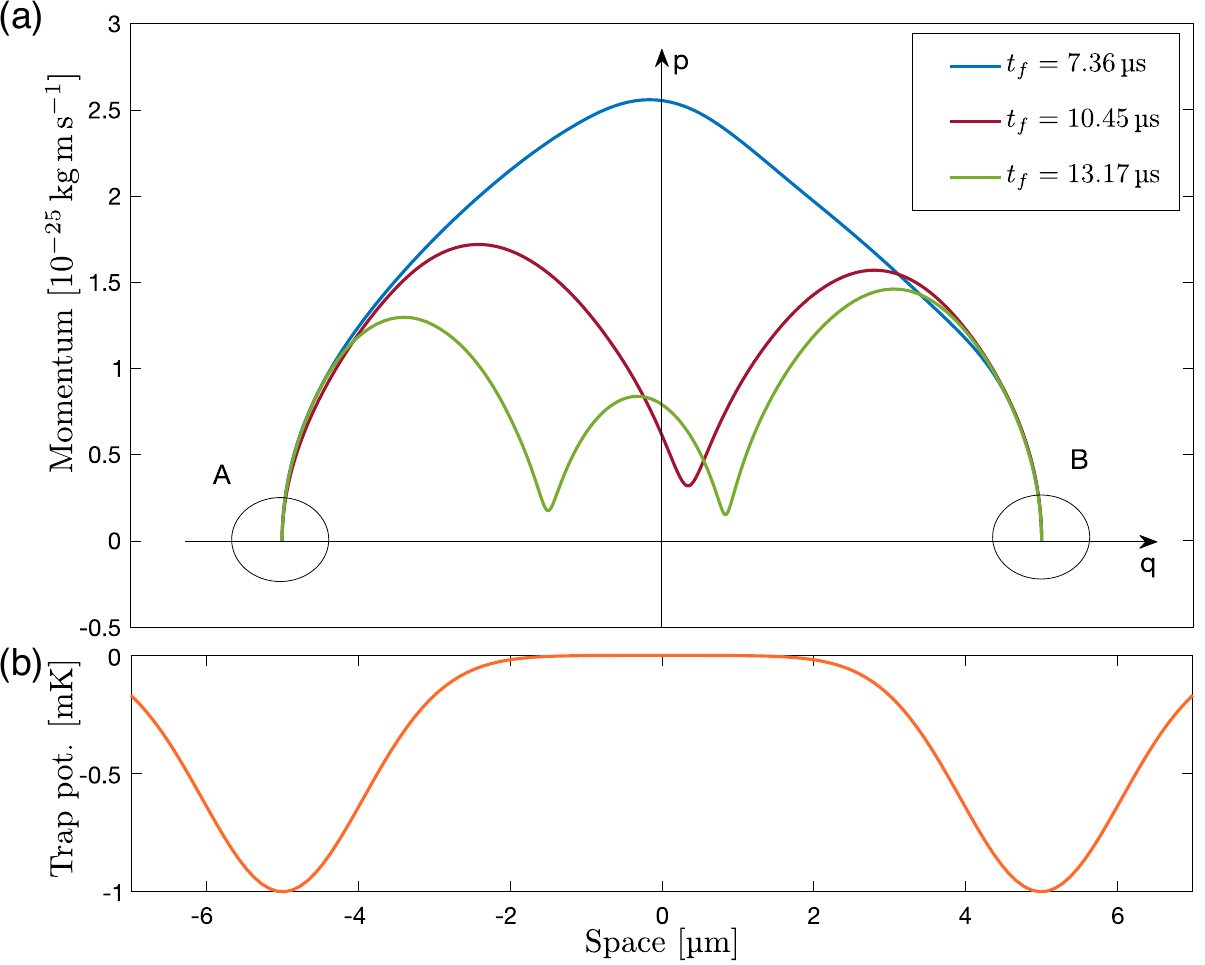} 
	\caption{Optimal control of the phase-space trajectories of the atoms driven by the tweezer field. We depict three distinct solutions of the optimality system corresponding to three final times $t_f$. (a) Phase-space trajectories corresponding to different optimal times: blue curve, $t_f=\qty{7.36}{\micro\second}$; red curve, $t_f=\qty{10.45}{\micro\second}$; and green curve, $t_f=\qty{13.17}{\micro\second}$. (b) Potential profile of the static initial and target traps.}
    \label{fig_Opt_traj_02}
\end{figure}

\begin{figure}[!ht]
	\centering
    \includegraphics[width=0.75\textwidth]{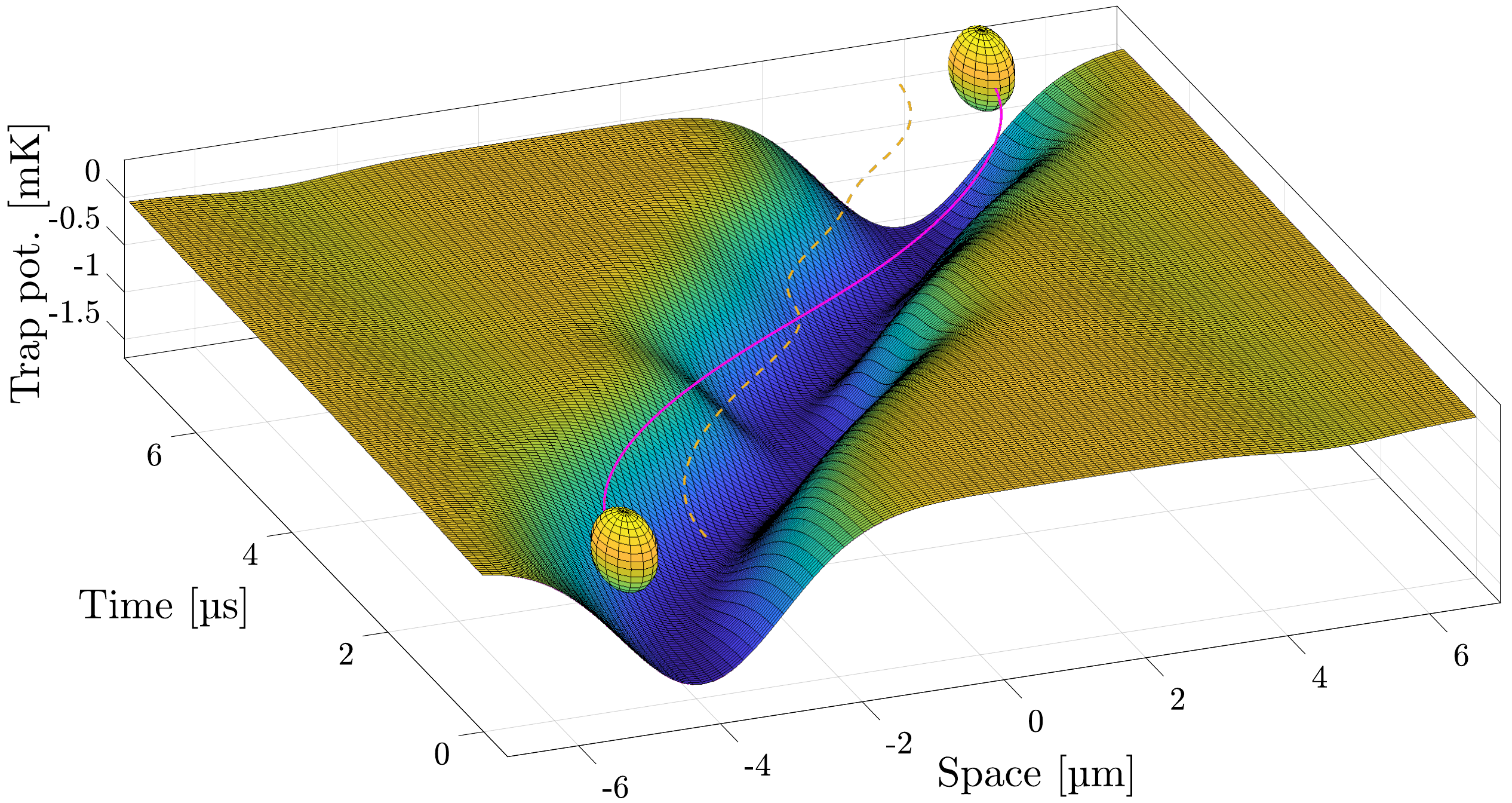}
    \caption{Time evolution of the atom position associated with the optimal time $t_f=\qty{7.36}{\micro\second}$ (blue curve in Fig.~\ref{fig_Opt_traj_02}(a)). The particle trajectory $x_{\text{opt}}(t)$ is depicted in magenta and the spheres indicate the initial and final positions of the atom. The 3D plot depicts the time evolution of the moving tweezer profile and, as a guide to eye, the yellow dashed curve depicts the evolution of the center of the tweezer's position. For this deterministic simulation the tweezer depth is fixed to $v=\qty{-1.5}{\milli\kelvin}$.}
    \label{fig_Opt_traj_time_01}
\end{figure}

\begin{figure}[!ht]
	\centering
	 \includegraphics[width=0.85\textwidth]{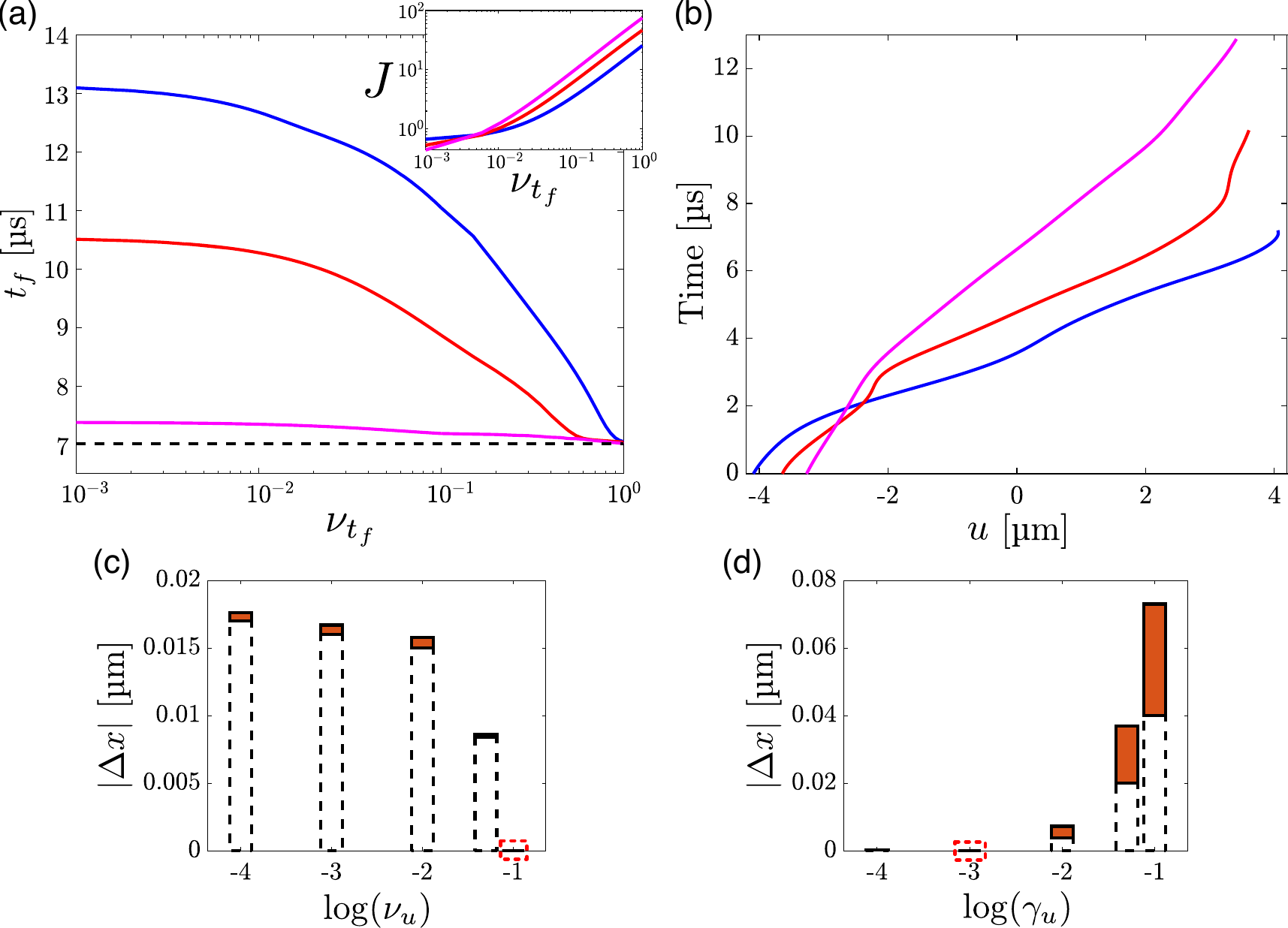}
	\caption{(a) Optimal time solutions as a function of the weight $\nu_{t_f}$. The dashed horizontal line indicates $t_{\text{lim}}$. The inset shows the estimation of the cost functional $J$ varying the weight $\nu_{t_f}$. (b) Controlled parameter $u$ as a function of time for the three optimal times. The colors refer to $t_f=\qty{7.36}{\micro\second}$ (blue), $t_f=\qty{10.45}{\micro\second}$ (red) and $t_f=\qty{13.17}{\micro\second}$ (magenta). (c) Maximum difference $\lvert \Delta x \rvert$ within $[0,t_f]$ occurring between the atom trajectories as a function of $\nu_u$ with $\log(\gamma_u)=-3$. Our optimal solution at $\log(\nu_u)=-1$ is highlighted by a red box. (d) Maximum difference $\lvert \Delta x \rvert$ within $[0,t_f]$ as a function of $\gamma_u$ with $\log(\nu_u)=-1$. Our optimal solution at $\log(\gamma_u)=-3$ is highlighted by a red box. Within each bar in (c) and (d), the orange area represents the range of variation of the optimal trajectory with respect to the reference case with $\log(\nu_u)=-1$ and $\log(\gamma_u)=-3$.}
    \label{fig_OptTime}
\end{figure}

In our simulations, we refer to the case of \isotope[88]{Sr} atoms~\cite{Cooper2018,Gavryusev2024} and the physical parameters are indicated in Table~\ref{tab_par}. The initial temperature of the trapped particle is $T=\qty{0.1}{\milli\kelvin}$. We set the atom temperature to this value to ensure reliable operation, deliberately overestimating it with respect to the typical temperature of around $\qty{0.01}{\milli\kelvin}$ at which atoms are prepared in tweezers~\cite{Hoelzl2023,Lu2024}. Reducing the temperature would further enhance the fidelity, although it is already remarkably high. The numerical results are shown in Figs.~\ref{fig_Opt_traj_02}-\ref{fig_OptTime}.
\begin{table}[!ht]
	\begin{center}
		\begin{tabular}{c|c|c|c} 
		  Trap distance $x_B-x_A$ ($\unit{\micro\meter}$)&
            Trap width $\zs_x$ ($\unit{\micro\meter}$)&
            Initial atom temperature $T (\unit{\milli\kelvin})$&
			Static trap depth $U_0$ $(\unit{\milli\kelvin})$\\ 
			\hline
			$\qty{10.0}{}$ & $\qty{1.5}{}$ & $\qty{0.1}{}$ & $\qty{1}{}$
		\end{tabular}\\[4mm] 
		\caption{Physical parameters used in our simulations. }\label{tab_par}
	\end{center}
\end{table}

Due to the nonlinearity, in general the optimality system may admit several solutions. In Fig.~\ref{fig_Opt_traj_02} we illustrate three possible solutions of the optimal control problem, classified by the time interval during which the control operates. We define the final time $t_f$ as the optimal time. We depict in Fig.~\ref{fig_Opt_traj_02}(a) the atom phase-space trajectory corresponding to the optimal times $t_f=\qty{7.36}{\micro\second}$ (blue curve), $t_f=\qty{10.45}{\micro\second}$ (red curve) and $t_f=\qty{13.17}{\micro\second}$ (green curve). The potential profiles of the static optical traps corresponding to the initial position $A$ of the atom on the left and to the target trap corresponding to the position $B$ are depicted by an orange curve in Fig.~\ref{fig_Opt_traj_02}(b). In order to illustrate the designed profile of the moving tweezer obtained by our optimal procedure, we depict in Fig.~\ref{fig_Opt_traj_time_01} the temporal evolution of the potential and of the atom position $x_{\text{opt}}(t)$ for the case $t_f=\qty{7.36}{\micro\second}$.  

The temporal profile of the tweezer parameters obtained by the optimal control solution related to the minimum value of the optimal time $t_{\text{lim}}$ can be understood by elementary considerations. In order to steer a particle over a distance $d$ in the minimum time, it is sufficient to ensure that the particle always undergoes the maximum acceleration available. The maximum value of the force associated with the tweezers field is $F_M= \pm v \sqrt{\frac{\zs}{2}}e^{-1/2}$, which corresponds to the acceleration $a_M=\frac{F_M}{m}$. The optimal control will increase the speed of the particle during the first half of the path by exerting a pulling force produced by positioning the moving tweezer center ahead of the desired trajectory, and decelerate it during the second part of the path always at the maximum rate $|a_M|$ via a pull-back force caused by lagging the potential behind the target path. This dynamic between tweezer and atom position is visualized in Fig.~\ref{fig_Opt_traj_time_01} by the yellow dashed and purple solid curves, respectively. It is of interest to note that these trajectories rarely overlap; moreover, the initial and final moving tweezer center coordinates do not coincide with those of the static traps. Finally, the moving potential has typically a depth much larger than the static ones in order to consider their presence as a small perturbation to the total potential landscape, ensuring that the atom is reliably driven in or out from the static traps. Assuming a constant depth of the moving tweezer $v=\qty{-1.5}{\milli\kelvin}$, we obtain the minimum theoretical transfer time of $t_{\text{lim}}=2\sqrt{\frac{d/2}{a_M}}=\qty{7.01}{\micro\second}$, which is in good agreement with the results of the simulations.

Our results are obtained by fixing the value of the parameter $v$ that describes the depth of the tweezer potential profile. The reason for this choice follows from the previous discussion. If we let $v$ be a free parameter to optimize, the problem of finding the minimum flying time may not be well posed. In fact, the final time can be made arbitrarily small by increasing the value of $v$ and consequently increasing the modulus of the force acting on the particle, which is physically restricted by the maximum available trapping power.


The definition of the cost associated with the process given in Eq.~\eqref{eq:cost} contains weights that constitute the free parameters of our model. Since the optimal solution is a minimizer (at least locally) of the cost functional and we require the optimal solution to satisfy various conditions (reaching the target position as close as possible, having a final velocity as small as possible, employing the minimum energy in the control), the weights can be adjusted in order to set the relative relevance of achieving a certain goal. In particular, $\nu_{t_f}$ is responsible for the weight attributed to completing the task in the minimum possible time interval. Since one of the challenges in the experiments is to maintain the atoms isolated from the environment, it is crucial to design optimal time control protocols. In our model, this can be achieved by increasing the value of $\nu_{t_f}$. In Fig.~\ref{fig_OptTime} we show the behavior of the optimal time with respect to the weight $\nu_{t_f}$. The plot shows that the optimal time decreases with increasing $\nu_{t_f}$ as expected, and the three solutions converge to the same value $t_{\text{lim}}$ indicated in Fig.~\ref{fig_OptTime}(a) by a dashed horizontal line. In the inset of Fig.~\ref{fig_OptTime}(a) we plot the behavior of the cost functional $J$ for varying weight $\nu_{t_f}$ for the three considered optimal times. In Fig.~\ref{fig_OptTime}(b) we plot the controlled parameter $u$ as a function of time for the three discussed optimal time solutions. For small values of $\nu_{t_f}$ the minimum $J$ corresponds to $t_f=\qty{13.17}{\micro\second}$. In this case, the main contribution to $J$ comes from the cost of the derivative of the controlled parameter $u$ (with $v$ kept constant), which is bigger for the blue line ($t_f=\qty{7.36}{\micro\second}$) than for the magenta one ($t_f=\qty{13.17}{\micro\second}$), as one can appreciate by looking at the inset of Fig.~\ref{fig_OptTime}(a). Increasing $\nu_{t_f}$, the contribution related to the optimal time $t_f$ becomes dominant, making the solution relative to the optimal time $t_f=\qty{7.36}{\micro\second}$ the global minimum of our optimality problem.

In our simulations we have considered the weights $\gamma_u=10^{-3}$ and $\nu_u=0.1$. This choice is arbitrary. Different values for $\gamma_u$ and $\nu_u$ lead to negligible deviations with respect to the trajectory $x_{\text{opt}}(t)$ illustrated by the magenta solid line in Fig.\ref{fig_Opt_traj_time_01}. To clarify this point, we depict the range of variation of the difference $\lvert \Delta x (t) \rvert = \lvert x(t)-x_{\text{opt}}(t) \rvert$ occurring within $[0,t_f]$. In Fig.~\ref{fig_OptTime}(c) we set $\gamma_u=10^{-3}$ and we vary $\nu_u$. In Fig.~\ref{fig_OptTime}(d) we set $\nu_u=0.1$ and we vary $\gamma_u$. The orange areas represent the range of variation of the optimal trajectory with respect to the reference case.

\section{Statistical ensemble} 
\label{sec:3}
In the classical framework, the dynamics of an ensemble of atoms in the presence of an external source of noise, represented by a thermal bath at the temperature $T_{\text{th}}$, is described by the Liouville Fokker-Planck (LFP) equation 
\begin{align}
	\dpp{f}{t}+ \frac{p}{m}\dpp{f}{x} - \dpp{U}{x} \dpp{f}{p}  -2\zg \dpp{(pf)}{p}- D_p \dsp{f}{p} - D_x \dsp{f}{x}  =0\;,\label{FP}
\end{align}
where $f$ denotes the atomic distribution function and $D_p, D_x$ and $\zg$ are diffusion coefficients subject to the fluctuation-dissipation relation $D_pD_x\geq \frac{\gamma^2}{4}$. 
The momentum diffusion coefficient is estimated as $D_p=\zg k_B T_{\text{th}}$, where $k_B$ is the Boltzmann constant~\cite{Isar1994,Manfredi2009}. The spatial diffusion coefficient is obtained by assuming the equality sign in the fluctuation-dissipation relation. 
The Fokker-Planck correction is an effective way to include various sources of experimental noise, including, e.g., laser intensity and frequency noise, trap position fluctuations, and heating by photon scattering~\cite{Labuhn2016,Bernien2017,Sheng2021,Bluvstein2022,Bluvstein2023}. We note that the standard Fokker-Planck formalism assumes uncorrelated white noise. In cases where the noise spectrum is structured or dominated by specific frequencies, i.e., colored noise, such as in certain parametric heating scenarios, this approach may not fully capture the system’s dynamics. The effective bath temperature should then be understood as a phenomenological parameter, and the LFP description has to be interpreted with care. The coefficient $\zg$,  measuring the coupling with the bath, does not have a direct experimental meaning itself, as the atom does not experience the effect of a true thermal bath. In combination with the effective bath temperature $T_{\text{th}}$ it determines the heating rate of the model, which can be adjusted to match the experimental measurable heating rate value. In the following, we keep $\zg=\qty{e-2}{\per\micro\second}$ fixed and change $T_{\text{th}}$ in order to show the performance of the optimal control under different heating rate scenarios.
	
The target region in the phase-space where the atom density should be found at the end of the process is identified by defining a non-negative function $f_{t_f}$ with a single maximum (typically one chooses an exponential localization). To measure the efficiency of the optimal transport, we consider the functional 
\begin{equation}
   \Phi'(f)=-\int_{\mathbb{R}^2} f_{t_f}(x,p)f(x,p,t_f)\dif x \dif p.
\end{equation}
Since the functions $f$ and $f_{t_f}$ are non-negative, the functional $\Phi$ is minimized if the solution $f$ at the final time concentrates around the maximum of the target function $f_{t_f}$. We formulate the ensemble optimal control problem as follows:
\begin{align}
	&\text{min}_{u,v} \; J(f,u,v)=\Phi'(f)+k(u,v) \label{OC_stat}\\
	&\text{s.t. } \; \textrm{Eq. }~\eqref{FP} \textrm{ holds true}. 
\end{align}
Here, in order to speed up the simulation and keep the computational cost low, we do not consider time as a controlled parameter. We use as a final time of our simulation the optimal time obtained before by solving the optimization problem on classical trajectories.
Similarly to the case of atoms along a deterministic trajectory, it is convenient to introduce the Lagrange functional 
\begin{align}
\label{eq:st_lagrangian}
\mathcal{L}\doteq&  J(f,u,v)\\
&+\int_0^{t_f} \int_{\mathbb{R}^2} \left(\dpp{f  }{t} +\frac{p}{m} \dpp{f}{x}   -\dpp{U}{x} \dpp{f}{p}
		-2\zg \dpp{(pf)}{p}- D_p \dsp{f}{p} - D_x \dsp{f}{x}  \right) h \dif x\dif p \dif t,
\end{align}
where the function $h$ represents the Lagrangian multiplier and is typically called an adjoint function. The set of equations corresponding to the solution of the optimal control problem is referred to as the optimality system and can be derived by imposing that the variation of $\mathcal{L}$ with respect to its arguments should vanish~\cite{Morandi2024}.
The resulting optimality system consists of the forward Liouville Fokker-Planck problem of Eq.~\eqref{FP}, along with a similar backward Liouville Fokker-Planck problem for the adjoint problem and the optimality condition 
\begin{align} \label{adl_FK}
        \dpp{h }{t} +\frac{p}{m} \dpp{h}{x} - \dpp{U}{x} \dpp{h}{p} -2\zg p\dpp{h}{p}+ D_p \dsp{h}{p} + D_x \dsp{h}{x}=0\quad  \mbox{in } \; [0,t_f]\times\R_{x} \times\R_{p},
\end{align}
with prescribed final value $\left.h\right|_{t=t_f} =f_
{t_f}(x,p) $ coinciding with the target atom distribution at the final time $t_f$. Finally, the control parameters are obtained by the equations
\begin{align} \label{cont_stat}
			\nu_{u}  \dst{u}{t}-	\zg_{u} u   =&  - \int_{\mathbb{R}^2}   \frac{\partial^2 U}{\partial x \partial u} h \dpp{f}{p}    \dif x  \dif p,\\
			\nu_{v}  \dst{v}{t}-	\zg_{v} v   =&  - \int_{\mathbb{R}^2}   \frac{\partial^2 U}{\partial x \partial v} h \dpp{f}{p}    \dif x  \dif p.
\end{align}
\begin{figure}[!ht]
	\centering
	\includegraphics[width=0.85\textwidth]{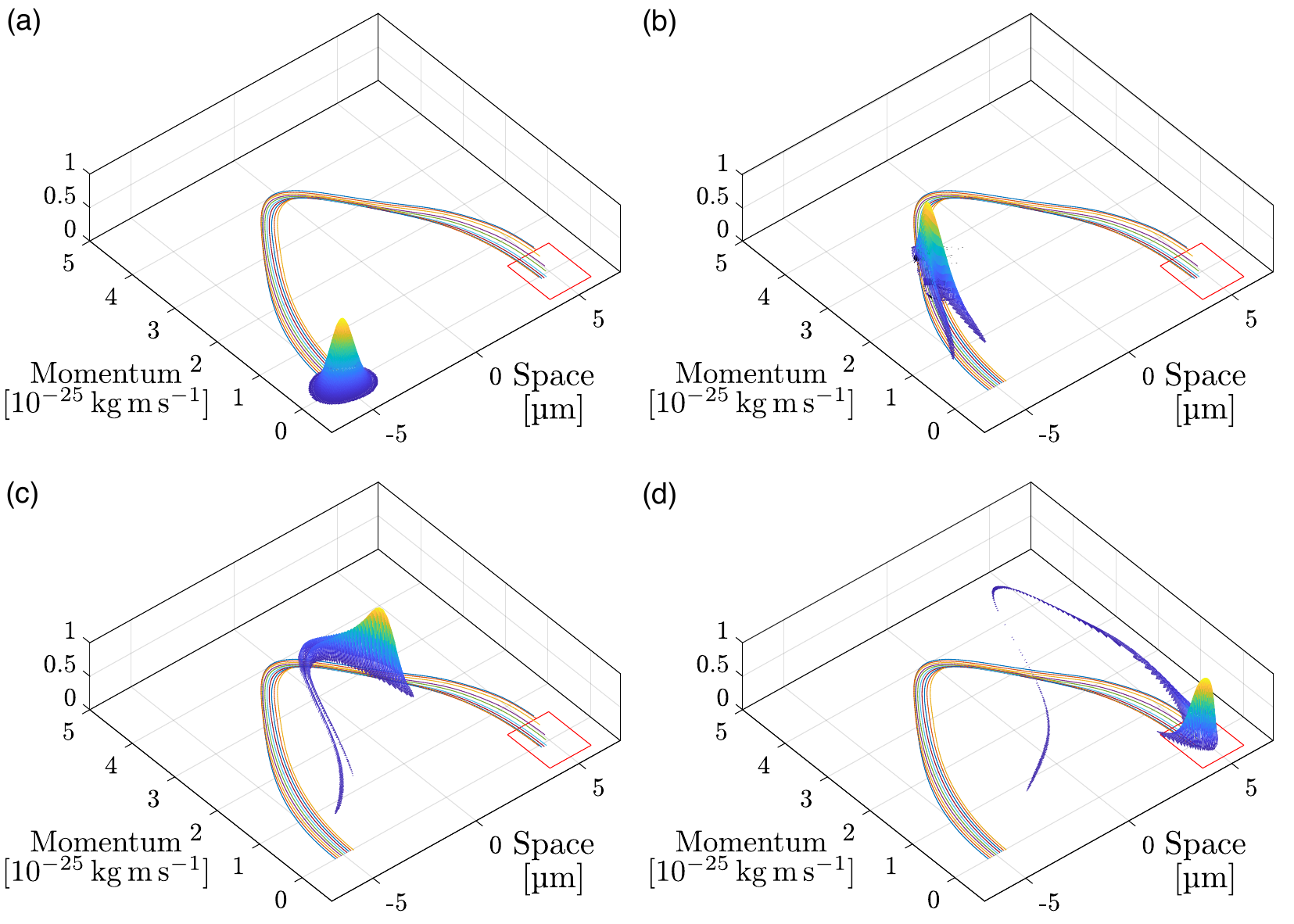}
    \caption{Optimal control procedure applied to the transfer of a well localized density of particles to the target region represented by the red rectangle. The plots are made at (a) $\qty{0.0}{\micro\second}$, (b) $\qty{2.4}{\micro\second}$, (c) $\qty{4.8}{\micro\second}$, and (d) $\qty{7.4}{\micro\second}$, considering a bath temperature of $T_{\text{th}}=\qty{0.1}{\milli\kelvin}$. The initial temperature of the atoms distribution is the same as indicated in Table~\ref{tab_par}.}
    \label{fig_cl_dens}
\end{figure}
%

Differently from the deterministic transport, here we will leave the moving tweezer depth $v$ as an optimizable quantity within the boundaries of the available laser power. In Fig.~\ref{fig_cl_dens} we depict the results of our optimal control procedure applied to an initial density with Gaussian localization in phase-space, representing atoms initially trapped inside a trap. The target region is represented by a red rectangle. We depict the solution at different times: $0,\,2.4,\,4.8,\,7.4\,\unit{\micro\second}$. The initial density is represented in Fig.~\ref{fig_cl_dens}(a). The bundle of trajectories in the phase-space represents the classical trajectories followed by the atom. Our simulation shows that the particle density is correctly driven into the target trap. 

\begin{figure}[!ht]
	\centering
    \includegraphics[width=0.6\textwidth]{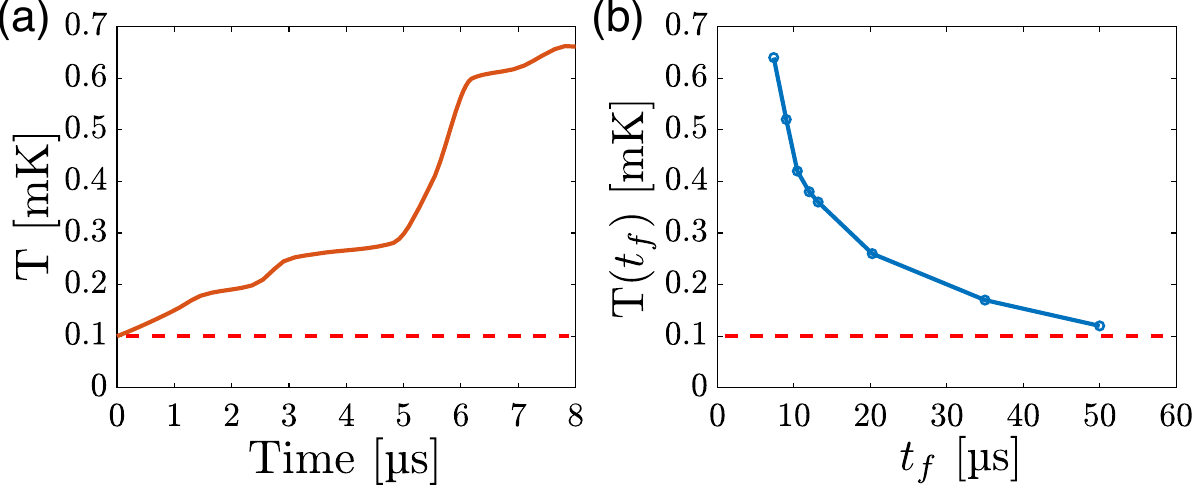}
	\caption{(a) Time evolution of the temperature $T$ of the atoms along the trajectory $x_{\text{opt}}(t)$ related to the optimal time $t_f=\qty{7.36}{\micro\second}$ (b) Final temperature $T(t_f)$ of the atoms as a function of the final time $t_f$. We extend our simulations to longer final times $t_f=9,12,20,35,50\,\unit{\micro\second}$, observing that $T(t_f)$ converges towards the initial temperature $T=\qty{0.1}{\milli\kelvin}$, represented by the red dashed line.}
    \label{fig_Temperature}
\end{figure}

In order to evaluate the reliability of the optimal control procedure against noise or external perturbations, we investigate the impact of the bath temperature in our simulations, with the expectation that the bath would lead to a heating of the atomic ensemble during the motion. The fidelity of the process is estimated by the percentage of the distribution function enclosed in the target area at the final time $t_f$ when the transporting tweezer potential is turned off, which is represented by the red rectangle centered on the target phase-space position depicted in Fig.~\ref{fig_cl_dens}. For simplicity, the fidelity of the process is estimated using a rectangle boundary with its edges at the typical trap extension $x=\pm\qty{1}{\micro\meter}$ and at the momentum $p_{td}=\pm\qty{0.63d-25}{\kilo\gram\meter\per\second}$ corresponding to the $\qty{1}{\milli\kelvin}$ static trap depth. Since the description of the system is fully in the classical regime, all atoms that reach the target trap with a momentum below $p_{td}$ remain trapped. Furthermore, the system does not thermalize in the final trap since there are no dissipation sources or heat exchange with an external bath. During the non adiabatic transfer process, the atom energy increases, as displayed in Fig.~\ref{fig_Temperature}(a), and we estimate that the final energy of the ensemble corresponds to a temperature around T$=\qty{0.64}{\milli\kelvin}$. We observe in Fig.~\ref{fig_Temperature}(b) that longer transport times allow for a smaller energy increase of the atoms, saturating towards the initial temperature of $\qty{0.1}{\milli\kelvin}$, which is consistent with the STA results~\cite{Hwang2025}. Consequently, our method enables us to reach a broad set of optimal solutions, ranging from fastest delivery to a desired compromise between transfer speed and energy increase and down to minimal heating.

\begin{figure}[!ht]
	\centering
    \includegraphics[width=0.75\textwidth]{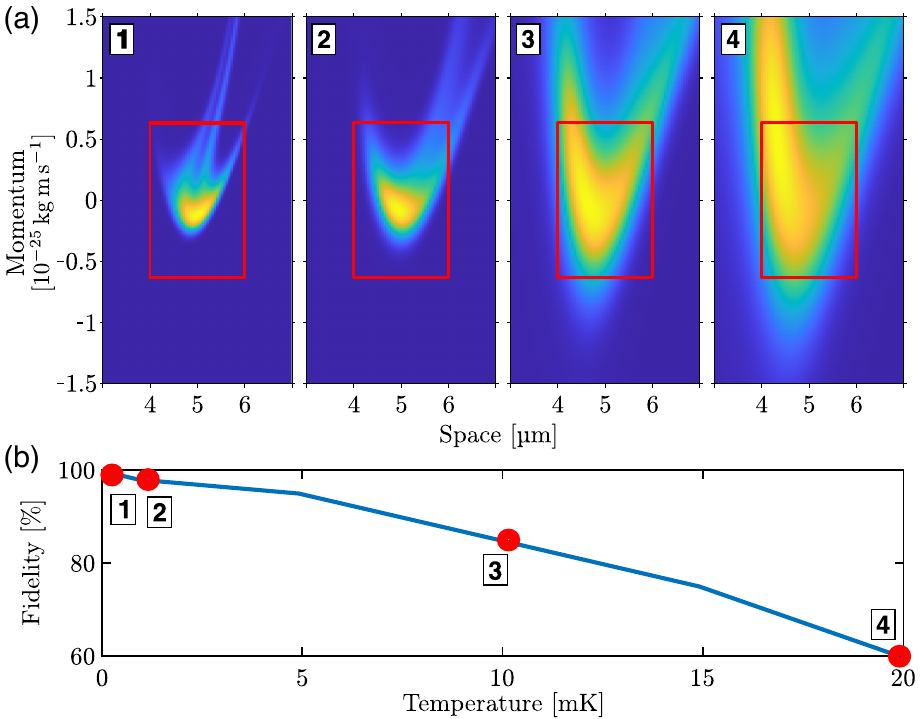}
	\caption{(a) Comparison of the atom distribution function at the final time $t_f$ at different bath temperatures $T_{\text{th}}=0.1,1.0,10.0,20.0\,\unit{\milli\kelvin}$. (b) Fidelity as a function of the bath temperature $T_{\text{th}}$. The numerical labels refer to the distribution functions in (a). The fidelity values at different bath temperatures are, in descending order, $99.97\%$, $98.75\%$, $85.64\%$ and $60.32\%$.}
    \label{fig_fidelity}
\end{figure}

In Fig.~\ref{fig_fidelity}(a) we compare the atom distribution at the final time $t_f$ by varying the bath temperature. For comparison, we use the same control potential as obtained as the optimal control for the bath temperature of $T_{\text{th}}=\qty{0.1}{\milli\kelvin}$. As expected, as the bath temperature increases, the distribution diffuses both in position and in momentum space and the control loses precision on steering the density inside the target trap. The plot of the fidelity degradation with the temperature is depicted in Fig.~\ref{fig_fidelity}(b). In our simulations, the maximum fidelity is found to be equal to $99.97\%$ for a bath temperature of $\qty{0.1}{\milli\kelvin}$. When the bath temperature increases to $\qty{20}{\milli\kelvin}$, the fidelity decreases to $60.32\%$. Intermediate cases are considered in Fig.~\ref{fig_fidelity}.

\begin{figure}[!ht]
	\centering
    \includegraphics[width=0.75\textwidth]{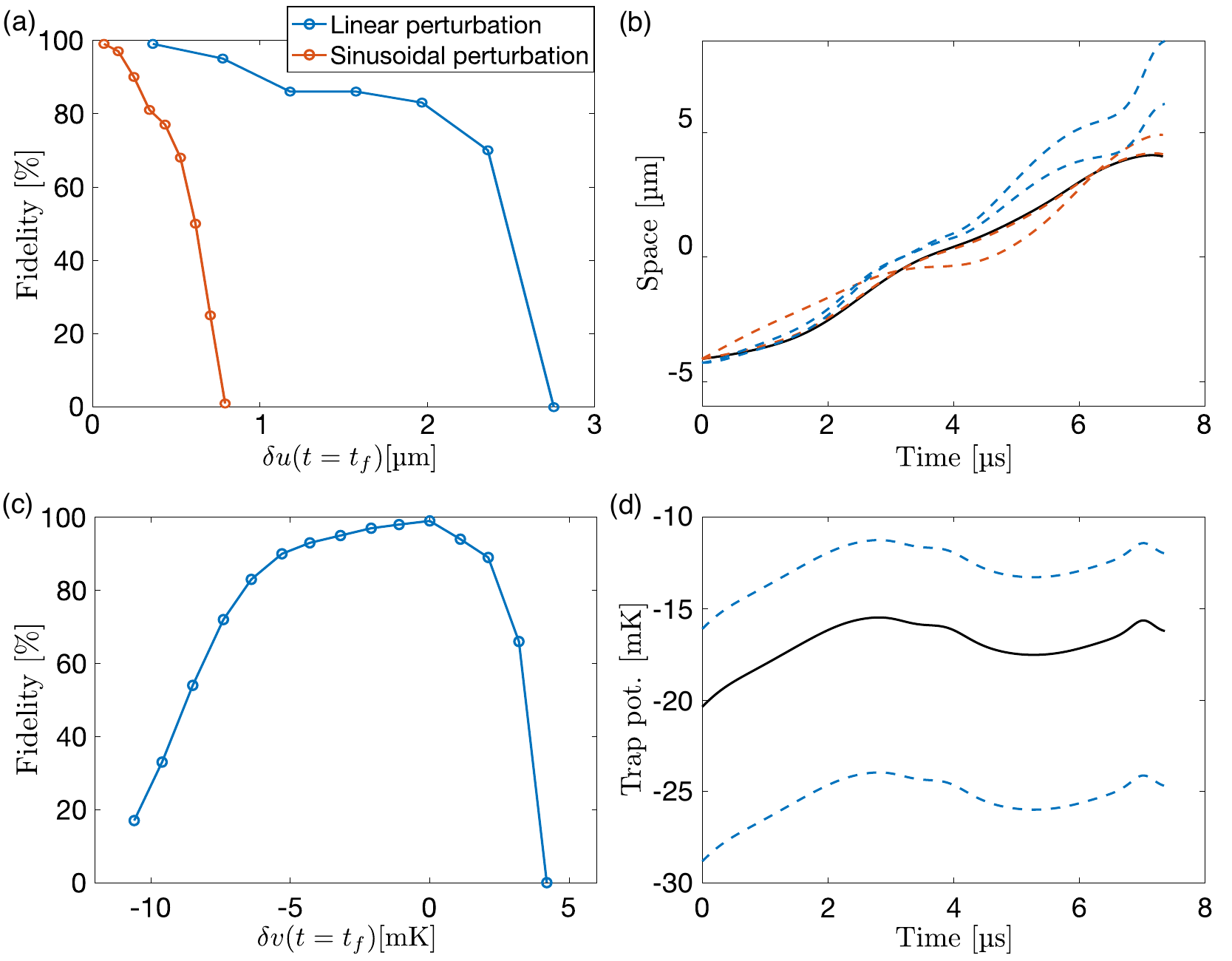}
	\caption{Perturbations of the parameters $u$ and $v$. (a) Variation of the fidelity corresponding to a perturbation of $u$ by a linear time-dependent signal (blue) and by a sinusoidal signal (red). The fidelity coefficient is plotted as a function of the variation on the final position of the tweezer ($t_f=\qty{7.36}{\micro\second}$). (b) Perturbed central position of the tweezer $u$. Red dashed curves refer to the sinusoidal signals and blue dashed curves refer to the linear perturbation. In both cases we plot the tweezer position corresponding to the minimum and to the maximum amplitude of the perturbation. (c) Fidelity coefficient as a function of the variation of the depth of the tweezer at the final configuration ($t_f=\qty{7.36}{\micro\second}$). (d) Constant in time perturbation of $v$; the two dashed lines mark the minimum and the maximum amplitude of the additive signal.}
    \label{fig_perturbations}
\end{figure}

To test the robustness of the optimization procedure, we perturbed the controlled parameters and evaluated the impact of the perturbation on the fidelity coefficient. The results are shown in Fig.~\ref{fig_perturbations}. We stress the parameter $u$ in two different ways. First, we add a linear time-dependent perturbation whose effect is to vary the velocity of the tweezer by a constant value. We plot the fidelity coefficient as a function of the variation of the final tweezer position with respect to the unperturbed one. Second, we add to $u$ a sinusoidal signal. The parameter $v$ is perturbed by increasing and reducing its value over time with a constant factor $\delta v$. In this case, the fidelity reaches a maximum of $99.97\%$ corresponding to the value $v=\qty{-16}{\milli\kelvin}$ obtained with our optimization procedure (see the control parameters in Figs.~\ref{fig_perturbations}(b)-\ref{fig_perturbations}(d)).

\section{Quantum description}
\label{sec:4}

In this section we solve the problem presented previously by a fully quantum approach. The Wigner description is a well established formalism introduced by Wigner in~\cite{Wigner1932} by which the statistical evolution of a quantum-mechanical system is represented in the classical phase-space. The Wigner formalism was successfully applied to model quantum dynamics in various contexts, e.g., for charged particles in semiconductors~\cite{Jacoboni2001,Morandi2012,Muscato2016,Camiola2020,Camiola2024}, graphene~\cite{Morandi2011}, quantum optics~\cite{Alonso2011,Weinbub2018}, control of entanglement~\cite{Morandi2022,Morandi2024a}, and particles in gravitational field~\cite{Manfredi2017}. The Wigner equation describing the evolution of a statistical ensemble of quantum particles reads
\begin{align} \label{Wig_eq}
 \dpp{f  }{t} +\frac{p}{m} \dpp{f}{x}   -\frac{1}{\ze} \Theta_{U}^\ze[  f ] -2\zg \dpp{(pf)}{p}- D_p \dsp{f}{p} - D_x \dsp{f}{x} =0 & \qquad \mbox{in } [0,t_f]\times\R_{x} \times\R_{p},
\end{align}
where we have introduced the pseudodifferential operator
\begin{align*}	
\Theta_{U}^\ze[ f ] 
\doteq& \frac{1}{2\pi i} \int_{\mathbb{R}^{2}} 
\left[ U  \left({x}+\frac{ \ze   \zh}{2}  \right)- U \left({x}-\frac{ \ze \zh }{2} \right)\right] f ({x},{p}',t)   e^{-i\left({p} -{p}'\right)  {\zh}}\dif {p}'  \dif  \zh.
\end{align*}

We have normalized the evolution equation by introducing the dimensionless parameter $\ze\doteq \frac{\hbar}{E_0 t_0}$, where $t_0$ is the characteristic time and $ E_0 $ is the characteristic energy of the atom. The parameter $\ze $ measures the quantumness of the atom dynamics. For a trapped particle $t_0$ is related to the trapping frequency $\omega$ (in a harmonic approximation of the trap potential) as $t_0 \sim \omega^{-1}$, which results in $\ze \sim \hbar \omega / E_0$, quantifying the importance of the discreteness of the quantum energy levels $\hbar \omega$ with respect to the particle energy $E_0$. For $\ze \ll 1$ we expect the quantum correction to the classical dynamics to be negligible. In particular, the well-known limit $ \lim_{\ze\rightarrow 0}\frac{1}{\ze}\Theta_{U}^\ze[ f ] = \dpp{U}{x}\dpp{f}{p}$ shows that the Wigner evolution equation coincides with the Liouville equation for vanishing $\ze$~\cite{Markowich1989}. Considering our previous examples of transport of Sr atoms inside traps of width $\qty{1.5}{\micro\meter}$, we obtain $\ze\simeq 10^{-3}$, which is sufficiently small to ensure that quantum effects may be neglected and justifies our previous choice to work with classical equations. In the case of lighter atoms trapped in smaller tweezers or within ultra-tight potentials~\cite{Wang2018,Tsui2020,Khazali2025}, quantum effects may become important. To test this scenario, we consider the limiting case of \isotope[6]{Li} atoms (approximately $15$ times lighter than \isotope[88]{Sr})~\cite{Serwane2011} and optical tweezers of width $\zs_x=\qty{0.3}{\micro\meter}$. In this case, the dimensionless parameter $\ze$ can be estimated to be equal to $\ze\simeq 0.22$. Furthermore, we fix the inter-trap distance to \qty{2}{\micro\meter} and we reduce the boundaries of the fidelity space edges to \qty{\pm0.2}{\micro\meter}.

The optimal control of atoms in the quantum regime proceeds similarly to the classical statistical case. The formulation of the optimal control problem is analogous to Eq.~\eqref{OC_stat}, where the LFP equation~\eqref{FP} is replaced by the Wigner equation~\eqref{Wig_eq}. We formulate the optimality problem~\cite{Morandi2024}
\begin{align}
	&\text{min}_{u,v} \; J(f,u,v)=\Phi'(f)+k(u,v) \label{OC_stat_q}\\
	&\text{s.t.} \; \textrm{Eq. }~\eqref{Wig_eq} \textrm{ holds true}, 
\end{align}
which  results in an analogous optimality system of~\eqref{OC_stat} where the adjoint function satisfies the equation
\begin{align} 
    \dpp{h }{t} +\frac{p}{m} \dpp{h}{x}  - \frac{1}{\ze} \Theta_{U}^\ze[  h ]-2\zg p\dpp{h}{p}+ D_p \dsp{h}{p} + D_x \dsp{h}{x} =0 \; \mbox{ in } \; [0,t_f]\times\R_{x} \times\R_{p},
\end{align}
with the final value condition $\left.h\right|_{t=t_f} =h_0$. The control parameters satisfy the equations
\begin{align}
    \nu_{u}  \dst{u}{t}-	\zg_{u} u   =  - \int_{\mathbb{R}^2}  h \Theta^\ze_{\dpp{U}{u}}  [f]    \dif x  \dif p \\
    \nu_{v}  \dst{v}{t}-	\zg_{v} v   =  - \int_{\mathbb{R}^2}  h \Theta^\ze_{\dpp{U}{v}}  [f]    \dif x  \dif p  \;.
\end{align}


\begin{figure}[!ht]
	\centering
    \includegraphics[width=0.75\textwidth]{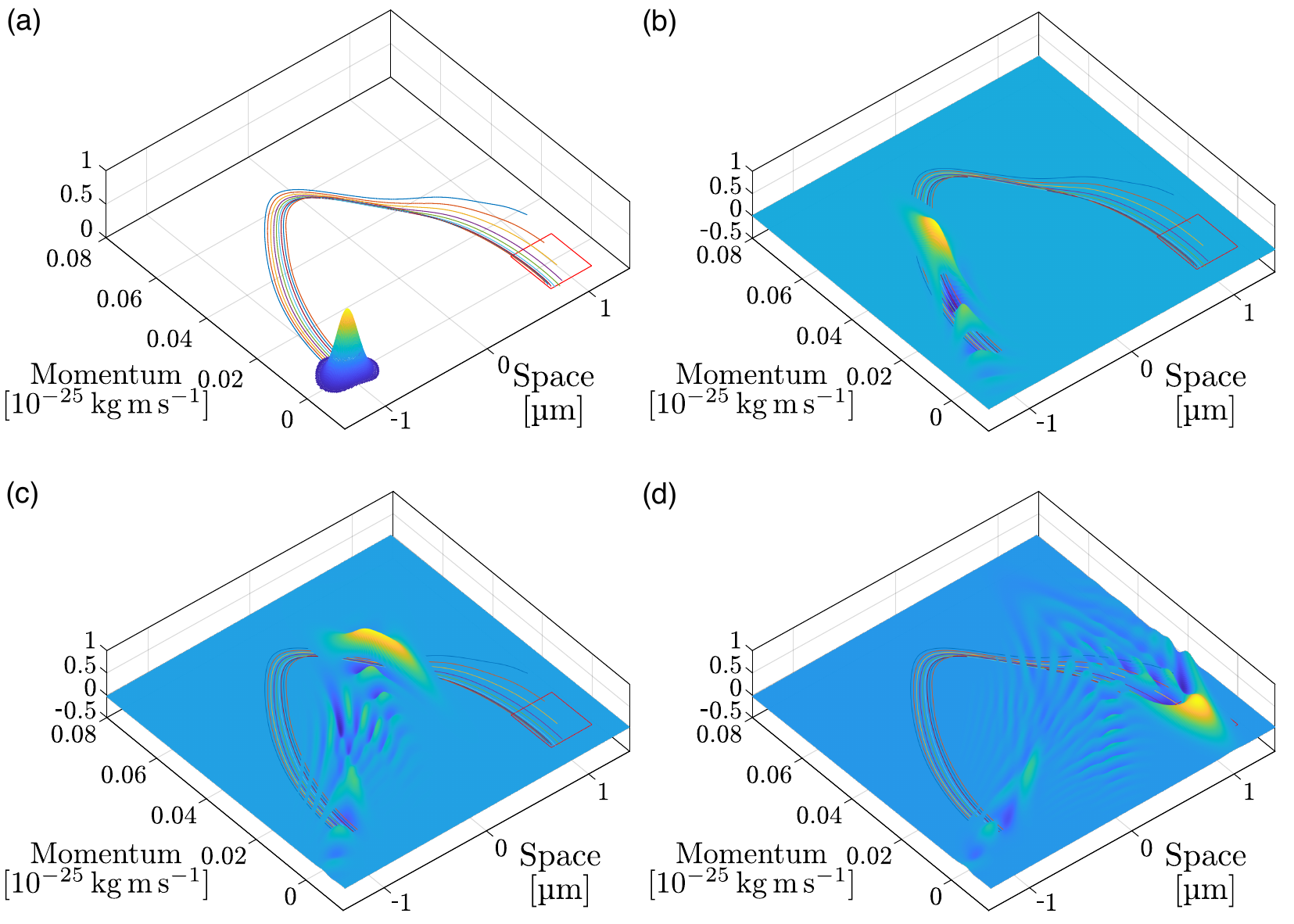}
	\caption{Optimal control of atomic transport in the case of quantum evolution. The Wigner function is plotted at times (a) $\qty{0.0}{\micro\second}$, (b) $\qty{2.4}{\micro\second}$, (c) $\qty{4.8}{\micro\second}$ and (d) $\qty{7.4}{\micro\second}$, with a bath temperature of $T_{\text{th}}=\qty{0.1}{\milli\kelvin}$. The temperature of the initial distribution is chosen as indicated in Table~\ref{tab_par}. For the sake of comparison, we depict the classical trajectories associated with the tweezer field.}
    \label{quantum sim}
\end{figure}

In Fig.~\ref{quantum sim} we show the results obtained by solving the optimality system. The quantum interference effects are clearly evident. We note that since we are considering atoms considerably lighter than in the previous case, the position and momentum scales are significantly modified. The Wigner approach allows us to describe the optimal control in the quantum regime along the same lines as in the classical case. The atom ensemble is described by a quasi-distribution that moves in the phase-space similarly to the classical one. As required by our optimization procedure, at the final time the Wigner distribution function concentrates at the target position in phase-space. For the sake of comparison, in the plot we depict the classical trajectories associated with the tweezer field. The fidelity coefficient is $98.95\%$, which is still remarkably high in this quantum case, demonstrating the effectiveness of our optimization procedure. The energy of the atoms increases from the initial temperature value of $\qty{0.1}{\milli\kelvin}$ to $\qty{0.78}{\milli\kelvin}$.
 

\section{Discussion}
\label{sec:5}
In this work, we have presented an optimal control procedure to design precise trajectories for atom transport, minimizing time and energy costs, while ensuring high fidelity. This has been achieved through careful modeling of noise and dynamics, described as stochastic effects. This approach differs from other methods employed in previous works. In~\cite{Hwang2023}, the authors rely on free-flight transport with acceleration and deceleration stages, which limits the precision of atom placement due to motional effects. In~\cite{Hwang2025}, the authors focus on providing a rapid, reliable transport mechanism using shortcuts to adiabaticity, allowing for fast atom movement with minimal mechanical heating, which is ideal for long-distance transport and to preserve the quantum state.
In~\cite{Chen2011} the authors explored optimal trajectories for atomic transport in harmonic traps using a combination of reverse engineering based on Lewis-Riesenfeld invariants and optimal control theory. They identified so-called bang-bang solutions for minimal time and bang-off-bang solutions for minimal displacement, with a focus on maintaining the alignment between the trap center and the center of mass within acceptable bounds. That work established foundational principles for time-optimal transport, but relied on fixed trajectories that may not fully account for experimental stochastic effects. Differently from that work, where trajectories are optimized in coordinate space, our method directly accounts for the full dynamics in phase-space, enabling precise control of the statistical distribution functions. Additionally, the incorporation of stochastic noise effects, modeled via Liouville Fokker-Planck and Wigner equations, allows us to simulate and mitigate experimental uncertainties, an aspect less emphasized in prior work.
In~\cite{Murphy2009} the authors addressed the transport of quantum states using a moving harmonic potential, while preserving encoded quantum information. They modeled scenarios with imperfect control inputs that affect the position of the potential well and demonstrated that perfect quantum information transfer is achievable in a nonadiabatic regime over any given distance. Their work provided important information on the interplay between control imperfections and quantum state fidelity, although without explicitly optimizing energy costs or time.
Our approach enables significantly faster atom transport compared to traditional adiabatic methods, achieving high fidelity even at reduced timescales. This is crucial for experiments requiring rapid initialization of large atom arrays, realized through rearrangement of a stochastically populated configuration. Adiabatic transport minimizes atom excitation, but requires extended timescales, which for arrays with more than a few hundred particles may compromise fidelity due to atom loss or heating~\cite{Pagano2024,Hwang2025}.

A key strength of our method lies in its robustness to experimental noise. By including stochastic terms (e.g., thermal bath effects) in our Liouville Fokker-Planck and Wigner models, we achieve reliable performance even in the presence of random perturbations such as laser fluctuations or environmental noise. As demonstrated in simulations with varying thermal bath conditions, our method actively incorporates stochastic effects and optimizes against them. The study in Ref.~\cite{Hwang2025} primarily addressed vibrational heating and atom loss, but did not directly account for other noise sources, focusing instead on speeding up adiabatic processes.

Another advantage of our approach is its flexibility. While the method in Ref.~\cite{Pagano2024} imposes constraints on pulse shapes (e.g., piecewise quadratic pulses), which limits adaptability, our procedure does not rely on a specific ansatz for the controlled parameters. This trait allows for the generation of control trajectories tailored to specific experimental setups. Shortucts to adiabaticity~\cite{Hwang2025} impose fewer constraints to preserve the classical initial state of a particle while allowing for transport over long distances. Our technique is also capable of minimally heating the atom by finding optimal solutions with $t_f\geq\qty{50}{\micro\second}$, which is consistent with STA results and only a factor of approximately $7$ longer than the fastest solution with $t_f=\qty{7.36}{\micro\second}$.

Finally, our method incorporates quantum corrections using the Wigner formalism, making it applicable to scenarios where quantum coherence is critical. In contrast, Refs.~\cite{Hwang2023,Hwang2025} assume classical dynamics, limiting their utility to those regimes in which quantum effects play a role. Preserving the motional quantum state of an atom may be valuable for experimental scenarios where quantum information is encoded in or coupled to the motional degree of freedom~\cite{Shaw2023}, or in quantum simulation platforms where quantum motion (e.g. quantum tunneling between traps and/or lattices sites~\cite{Dutta2015,Schaefer2020}) is important.  The quantum transport framework of Ref.~\cite{Murphy2009} aligns with our extension into the quantum regime, where we employed the Wigner formalism to incorporate quantum corrections. However, our method’s inclusion of stochastic effects and cost functionals tailored for common experimental scenarios ensures robustness against perturbations such as thermal noise and trap imperfections. Our quantum Wigner method can potentially be applied in these scenarios by imposing the constraint on the preservation of the initial motional quantum state.

\section{Conclusion}
\label{sec:6}
The optical control of ultracold atoms is essential to many quantum science and technology applications, e.g. for the realization of atom-based quantum simulators and computers with single-atom control. In this framework, the necessity of flexible and efficient schemes to steer the position of atoms trapped in optical tweezers arises. We have proposed a technique based on solving an optimal control problem applied to the transport equation, with the requirement to ensure high transport fidelity with minimal time and energy costs. 

We have modeled the transport process first with a classical Liouville equation and then through a quantum Wigner equation in order to investigate the presence of quantum features. The classical trajectories, computed by solving an optimal problem for the Hamiltonian equations of a single particle moving in a one-dimensional direction, have been used as an initial guess for the optimization of the transport problem at a statistical level. The control procedure acts on a moving tweezer potential, assumed to have a Gaussian shape, by modulating its depth and center coordinates. Furthermore, the transport time is considered as a free parameter to be minimized. Such a choice is broadly applicable in experimental setups where the tweezers result from an SLM or an AOD that control and shape the optical traps. The analysis is carried out assuming typical experimental parameters and uses the mass of \isotope[88]{Sr} atoms in the classical case and of \isotope[6]{Li} atoms in the quantum case, but it is easily adaptable for other atomic species. 

We have studied the optimal control procedure to steer an atom from an initial trap to a target site distanced by $\qty{10}{\micro\meter}$, aiming to reach a fidelity above $99.5\%$. We have identified an absolute minimum with a flying time of $\qty{7.36}{\micro\second}$, which is close to the theoretical lower physical boundary, and with a high fidelity of $99.97\%$ in the classical case. Such fidelity would enable us to rearrange 100 atoms with a total success rate of $97\%$ which is consistent with current state-of-the-art implementations~\cite{Bluvstein2023,Pichard2024,Manetsch2024,Gyger2024} but on a timescale of only $\qty{1}{\milli\second}$.
The optimum solution derived in the quantum case, leveraging the same trajectory and flying time, provides a fidelity of $98.95\%$. We have tested the robustness of the result by modifying the weight of the time in the cost functional and we have identified other local-minimum optimal solutions. Additionally, we have checked the robustness of the achieved fidelity against significant fluctuations of trap depth and position. Furthermore, both the classical and the quantum transport equations integrate Fokker-Planck terms to take into account the effect of the perturbations that are present in typical experiments. By varying the temperature of the external bath, we have quantified the effect of the noise in decreasing the fidelity of reaching the target state.

Finally, we have highlighted the strengths and limitations of our approach relative to the state of the art for inter-trap atom transport. Our results provide a fast non adiabatic method for relocating atoms from an initial configuration to a desired target arrangement, minimizing time and energy costs while ensuring high fidelity. This can be highly valuable in quantum simulation or computation experiments that require the initialization of large single-atom arrays.

\appendix*
\section{Derivation of optimality conditions}\label{app:A}
In this appendix we derive the optimality conditions corresponding to the optimal transport of the atoms. \\
{\it Deterministic transport.} We start with the case of a single atom in the classical framework. 
First, we compute the derivatives with respect to the variables of the adjoint problem:
    \begin{align*}
		(\nabla_x^h \mathcal{L},\delta x^h) &=\int_0^{t_f}\left[  \dot{p}+\frac{\partial U}{\partial x}\right]\delta x^h \dif t,\\
		(\nabla_p^h \mathcal{L},\delta p^h) &=\int_0^{t_f}\left[  \dot{x}-\frac{p}{m}\right]\delta p^h \dif t.
	\end{align*}
Imposing that $(\nabla_x^h \mathcal{L},\delta x^h)=0$ and $(\nabla_p^h \mathcal{L},\delta p^h)=0$, we obtain the Hamiltonian equations~\eqref{Ham_traj}. In a similar way, we obtain the equations for the adjoint variables
    \begin{align*}
		(\nabla_x \mathcal{L},\delta x) &=\int_0^{t_f}\left[  {\dot{p}}^h-\frac{\partial^2 U}{\partial x^2}\right]\delta x \dif t+\left[p^h (t_f) + \nu_x (x(t_f)-x_B)\right]\zd x ({t_f})\\
		(\nabla_p \mathcal{L},\delta p) &=\int_0^{t_f}\left[  -{\dot{x}}^h-\frac{{\dot{p}}^h}{m}\right]\delta p \dif t +\left[x^h(t_f) + \nu_x p(t_f)\right]\zd p ({t_f}).
	\end{align*}
We proceed by calculating the G\^{a}teaux derivatives of the Lagrangian~\eqref{cl_lagrangian} with respect to the control parameters.  We obtain
	\begin{align*}
		(\nabla_u \mathcal{L},\delta u) &=\left[ - \int_0^{t_f}    \frac{\partial^2 U}{\partial x \partial u}\;  x^h \dif t  + \zg_u u- \zn_u \dst{u}{t} \right]\delta u\\
		(\nabla_v \mathcal{L},\delta v) &=\left[ -\int_0^{t_f}  \frac{\partial^2 U}{\partial x \partial v} \; x^h \dif t  + \zg_v v- \zn_v \dst{v}{t} \right]\delta v\\
		(\nabla_{t_f} \mathcal{L},\delta t_f) &=\nu_{t_f}t_f + \biggl(\frac{\nu_x}{m}(x_B-x(t_f))-\nu_p\left.\frac{\partial U}{\partial x}\right|_{t=t_f}\biggr)p(t_f)+\frac{\gamma_u}{2}u^2(t_f)+\frac{\gamma_v}{2}v^2(t_f)\;,
	\end{align*}
where we have assumed that the controls satisfied the conditions $(\dpt{u}{t},\dpt{v}{t})(0)=(0,0)$ and $(\dpt{u}{t},\dpt{v}{t})(t_f)=(0,0)$. \\ 
{\it Statistical ensemble.} 
We compute the G\^{a}teaux derivative of the Lagrangian functional~\eqref{eq:st_lagrangian} with respect to its arguments. First, the derivative of $\mathcal{L}$ with respect to $h$ reads
	\begin{align*}
		(\nabla_h\mathcal{L}, \delta h)= \int_0^{t_f}\int_{\mathbb{R}^2} \biggl(&\frac{\partial f}{\partial t}+\frac{p}{m} \frac{\partial f}{\partial x} - \dpp{U}{x} \dpp{f}{p}  \\
		& -2\zg \dpp{(pf)}{p}- D_p \dsp{f}{p} - D_x \dsp{f}{x}\biggr)  \delta h \dif x \dif p \dif t .
	\end{align*}
	The equation $(\nabla_h\mathcal{L},\delta h)=0$ corresponds to the weak formulation of the Liouville-Fokker-Planck equation. The variation of the Lagrangian functional $\mathcal{L}$ with respect to $f$ gives
	\begin{align*}
		(\nabla_f\mathcal{L},\delta f)=& \int_0^{t_f}\int_{\mathbb{R}^2} \biggl(-\frac{\partial h}{\partial t}-\frac{p}{m} \frac{\partial h}{\partial x} + \dpp{U}{x} \dpp{h}{p}  \\
		& -2\zg \dpp{(ph)}{p}+ D_p \dsp{h}{p} + D_x \dsp{h}{x}\biggr) \delta f \dif x \dif p \dif t \\
		&+\int_{\mathbb{R}^2} [h(x,p,T)-f_{t_f}(x,p)]\delta f(x,p,t_f)\dif x \dif p,
	\end{align*}
	where we used  $\delta f(x,p,0)=0$. The stationarity condition  $(\nabla_f\mathcal{L},\delta f)=0$ leads to the adjoint equation 
	\begin{equation*}
		\begin{cases}
			&\displaystyle \frac{\partial h}{\partial t}+\frac{p}{m}\frac{\partial h}{\partial x}-\dpp{U}{x} \dpp{h}{p} -2\zg \dpp{(ph)}{p}- D_p \dsp{h}{p} - D_x \dsp{h}{x}=0,\\
			&\displaystyle h|_{t=t_f}=f_{t_f}
		\end{cases}.
	\end{equation*}
The derivation of the optimality conditions for $u$ and $v$ follows a procedure similar to that of classical trajectories.\\
{\it Quantum description.} 
The procedure to obtain this optimality system is similar to the previous cases for the quantum Lagrangian functional 
\begin{align*}
\mathcal{L}=& J+\int_0^{t_f} \int_{\mathbb{R}^2}\left(\dpp{f  }{t} +\frac{p}{m} \dpp{f}{x}   -\frac{1}{\ze} \Theta_{U}^\ze[  f ] -2\zg \dpp{(pf)}{p}- D_p \dsp{f}{p} - D_x \dsp{f}{x}\right) h \dif x\dif p\dif t
\end{align*}
The derivation of the optimality conditions for the quantum Lagrangian functional is discussed in detail in~\cite{Morandi2024}.

\begin{acknowledgments}
We acknowledge insightful discussions with Dr. Lorenzo Buffoni. The work was developed under the auspices of GNFM (INdAM). This project received funding from Consiglio Nazionale delle Ricerche PASQUA Infrastructure, QuantERA ERA-NET Cofund in Quantum Technologies project MENTA, from the Italian Ministry of Education and Research PRIN 2022SJCKAH HIGHEST, and, in the context of the National Recovery and Resilience Plan and Next Generation EU, from Project No. PE0000023-NQSTI and from M4C2 investment 1.2 project MicroSpinEnergy (V.G.).
\end{acknowledgments}

\bibliography{apssamp}
\bibliographystyle{apsrev4-2}

\end{document}